# Universal Laws of Human Society's Income Distribution[*]

Yong Tao[†]

College of Economics and Management, Southwest University, Chongqing, China

**Abstract:** General equilibrium equations in economics play the same role with many-body Newtonian equations in physics. Accordingly, each solution of the general equilibrium equations can be regarded as a possible microstate of the economic system. Since Arrow's Impossibility Theorem and Rawls' principle of social fairness will provide a powerful support for the hypothesis of equal probability, then the principle of maximum entropy is available in a just and equilibrium economy so that an income distribution will occur spontaneously (with the largest probability). Remarkably, some scholars have observed such an income distribution in some democratic countries, e.g. USA. This result implies that the hypothesis of equal probability may be only suitable for some "fair" systems (economic or physical systems). From this meaning, the non-equilibrium systems may be "unfair" so that the hypothesis of equal probability is unavailable.

## 1. Introduction

Statistical physics was extraordinarily successful in dealing with the physical properties of systems which consist of huge numbers of particles. Although the number of economic agents in an economic system is less than the counterpart quantity in physical system, there are still $10^6$ firms and $10^7$ households, and these are large numbers [1]. Because of this, the statistical physics was also applied in studying the statistical properties of complex economic systems consisting of a large number of economic agents [1-12]. It was widely accepted that [6] the data analysis of empirical distributions of income reveals a two-class distribution. As is well known,

[*] Project supported by the Fundamental Research Funds for the Central Universities, (Grant No. SWU1409444); The National Social Science Foundation of Great Invite Public Bidding Project (Grant No. 11&ZD047); The National Soft Science Research Plan (Grant No. 2013GXS4D143); The Fundamental Research Funds for the Central Universities, (Grant No. SWU1309369).
[†] Corresponding author.
   E-mail address: taoyingyong@yahoo.com

Yakovenko [6] has used the income data from 1983-2000 in USA to confirm that the majority of the population (lower class) obey Boltzmann distribution (i.e., exponential distribution) and the small fraction of the population (upper class) obey Pareto distribution (i.e., power-law); see Figure 6 in Ref [6].

Nonetheless, recently, Kusmartsev et al [8-9] used the income data from 1996-2008 in USA to find that the majority of the population (lower class) abnormally obey Bose-Einstein distribution but the small fraction of population (upper class) still obey Pareto distribution; see Figure 1 in Ref [8]. Thus, Yakovenko's and Kusmartsev's studies seemed to contradict the agreement that the lower class of population will obey the universal law. However, when Tao [13] applied the principle of maximum entropy into Arrow-Debreu's general equilibrium model, he found that the equilibrium income distribution among social members will obey Boltzmann distribution for a moderately competitive society but will obey Bose-Einstein distribution for an extremely competitive society. In particular, the economic crisis (e.g. financial crisis in 2008) can be just regarded as a Bose-Einstein condensation among social members [7]. Thus, Tao's results not only explain why Yakovenko's and Kusmartsev's research results are different in investigating the lower class of population, but also further imply that the lower class of population well occupy the general equilibrium state, and that the upper class may be far from the general equilibrium. From this meaning, Tao indeed provided an alternative theoretical explanation to clarify the universal law of humans' income distribution. On the basis of Tao's previous work [13], this paper will translate the standard procedure of statistical physics into the standard language of neoclassical economics. We hope that such an attempt can serve some communications between physicists and economists.

The main purpose of statistical physics is to obtain the most probable macrostate which will contain the most microstates. In general, each microstate is specified by a collection of generalized coordinates and generalized momentums:

$$(q_1,...,q_D; p_1,...,p_D), \qquad (1)$$

where $q_i$ and $p_i$ denote the generalized coordinate and the generalized momentum of the $i$ th particle respectively.

Since each particle must obey the Newtonian equations, each $q_i$ and $p_i$ will be determined by the Hamiltonian equations[1], which are on the basis of the Principle of Least Action. As such, we immediately comprehend that each microstate is essentially a solution of the $D$-body Newtonian equations at a given time. To obtain the most probable macrostate, the key step is to assume that every microstate will occur with an equal probability.

Similar to the physical system in which the behaviors of particles must obey the Newtonian equations, the optimal behaviors of social members in the economic system will be governed by Arrow-Debreu's general equilibrium equations (ADGEE) which are the cornerstone of neoclassical economics [14]. In neoclassical economics, ADGEE is derived by Rational Person Hypothesis, which is the fundamental economic principle. It is here interesting to point out that the Principle of Least Action in physics is also thought of as Nature's own principle of economy.

---

[1] That is, $\begin{cases} \dfrac{dq_i}{dt} = \dfrac{\partial H}{\partial p_i} \\ \dfrac{dp_i}{dt} = -\dfrac{\partial H}{\partial q_i} \end{cases}$ for $i = 1,...,D$.

The solution to ADGEE is called the competitive equilibrium which corresponds to an optimal income allocation. Similar to many-body Newtonian equations, ADGEE will have multiple solutions (for a long time). Then by Arrow's Impossibility Theorem social members will face the "dilemma of social choice". This dilemma implies that the best income allocation cannot be found. Interestingly, this paper shows that the principle of maximum entropy will provide a possible way of removing the "dilemma of social choice" in a just society which obeys Rawls' fairness. Later, we shall see that each solution to ADGEE can be regarded as a possible microstate of the economic system, so that the principle of maximum entropy may be available in a society which captures Rawls' fairness.

## 2. Arrow-Debreu's General Equilibrium Model

Following the standard framework of neoclassical economics [14], we assume that there are $N$ consumers, $N$ firms and $L$ types of commodities. Every consumer $i = 1,...,N$ is specified by a consumption set $X_i \subset R^L$, a preference relation $\succsim_i$ on $X_i$, an initial endowment vector $\omega_i \in R^L$. Each firm $j = 1,...,N$ is characterized by a production set $Y_j \subset R^L$. We denote by $x_i = (x_{1i},..., x_{Li})$ the consumption vector of the $i$ th consumer, where $x_i \in X_i$ and $x_{ki} \geq 0$ for $k = 1,...,L$. We denote by $y_j = (y_{1j},..., y_{Lj})$ the production vector of the $j$ th firm, where $y_j \in Y_j$. Adopting the input-output model's convention [15], $y_j$'s positive component denotes output and negative component denotes input. Without loss of generality, we assume that all the firms only produce the $m$ th type of commodity, namely $y_{mj} \geq 0$ for $j = 1,...,N$ and $y_{lj} \leq 0$ for $l \neq m$. The economic meaning of this assumption is that there is one industry only. Such an assumption will help us to simplify the complexity of calculation, and does not influence our final result; for details see Tao's [13] discussion about multiple industries.

Because our main purpose[2] is to investigate income allocation and income distribution among social members, we further assume that the $i$ th consumer is the owner of the $i$ th firm, where $i = 1,...,N$. Thus, the revenue of the $i$ th firm is identified with the income of the $i$ th consumer[3]. It is carefully noted that "income" in this paper is somewhat different from "wealth". For instance, "wealth" may be due to either property inheritance or individual effort, but "income" in this paper only involves individual effort.

---

[2] Since national income differences may be the biggest problem facing the world today [16], our attention will be concentrated on the income distribution among social members.
[3] From the empirical point of view, these entrepreneurs can be regarded as a sampling about income allocation.

Similar to the physical system in which the behaviors of $D$ particles, $(q_1,...,q_D; p_1,...,p_D)$, must obey Newtonian equations, the optimal behaviors of $N$ consumers, $(x_1^c,...,x_N^c; y_1^c,...,y_N^c)$, must obey the following ADGEE[4] [14]:

(a). Profit maximization: For every firm $i$, $y_i^c \in Y_i$ maximizes profits in $Y_i$; that is,

$p \cdot y_i \leq p \cdot y_i^c$ for all $y_i \in Y_i$

(b). Utility maximization: For every consumer $i$, $x_i^c \in X_i$ is maximal for $\succsim_i$ in the budget set:

$$\left\{ x_i \in X_i : p \cdot x_i \leq p \cdot \omega_i + \sum_{j=1}^{N} \theta_{ij} p \cdot y_j^c \right\}.$$

(c). Market clearing: $\sum_{i=1}^{N} x_i^c = \sum_{i=1}^{N} \omega_i + \sum_{i=1}^{N} y_i^c$.

Adopting the technical term in the neoclassical economics, here we say that $(x_1^c,...,x_N^c; y_1^c,...,y_N^c)$ and the price vector $p = (p_1,..., p_L)$ will constitute a general equilibrium (or competitive equilibrium). Because the general equilibrium exhibits the optimal behavior of each consumer, and also because these optimal behaviors are the most specific descriptions one can get, the general equilibrium $(x_1^c,...,x_N^c; y_1^c,...,y_N^c)$, identified with the role of (1) in statistical physics, can be regarded as a microstate of the economic system. From this meaning, the general equilibrium in economic system plays the same role with the microstate in physical system.

Nevertheless, it is worth mentioning that the ADGEE (a)-(c) do not describe the long-run behaviors of consumers. This is because the long-run level of profits for a competitive firm (consumer) is a zero level of profit [7,17]. That is to say, we must have:

$p \cdot y_i^c = 0$ (2)

for $i = 1,..., N$.

By substituting (2) into (a)-(c), Tao [13] proposed the following long-run ADGEE:

(d). For every firm $i$, there exists $y_i^* \in Y_i$ such that $p \cdot y_i \leq p \cdot y_i^* = 0$ for all $y_i \in Y_i$.

---

[4] Newton's equations and Arrow-Debreu's general equilibrium equations do not share the same structure. The only similarity between them is that they determine optimal behaviors: Newtonian equations determine the optimal behaviors of particles; likewise, Arrow-Debreu's general equilibrium equations determine the optimal behaviors of social members.

(e). For every consumer $i$, $x_i^* \in X_i$ is maximal for $\succsim_i$ in the budget set:

$$\{x_i \in X_i : p \cdot x_i \leq p \cdot \omega_i\}.$$

(f). $\sum_{i=1}^{N} x_i^* = \sum_{i=1}^{N} \omega_i + \sum_{i=1}^{N} y_i^*$.

Similar to the ergodic hypothesis in the statistical physics, we believe that the statistical regularities of economic system shall emerge over a large time scale. Therefore, we will mainly focus on the long-run ADGEE (d)-(f) which describe the long-run "interactions" between consumers (social members).

## 3. Long-Run Equilibrium Solutions

Tao [13] has proved that the equations (d)-(f) will have multiple solutions[5] (or long-run equilibria):

$$\left(x_1^*, \ldots, x_N^*; t_1 z, \ldots, t_N z\right), \qquad (3)$$

where, $x_i^*$ for $i = 1, \ldots, N$ and $z = (z_1, \ldots, z_L)$ are fixed vectors, and $\{t_i\}_{i=1}^{N}$ satisfies:

$$\begin{cases} t_i \geq 0 \quad for \quad i = 1, 2, \ldots, N \\ \qquad \sum_{i=1}^{N} t_i = 1 \end{cases}. \qquad (4)$$

Also, $z$ obeys

$$p \cdot z = 0, \qquad (5)$$

where $p = (p_1, \ldots, p_L)$ stands for the equilibrium price vector.

The equation (5) guarantees that every firm (consumer) in the long-run equilibria only obtains zero economic profit. Then by (3), each firm $i$ will obtain $t_i p_m z_m$ units of revenue [13], where $z_m$ denotes the $m$ th component of $z$. Because the consumer $i$ is the owner of the firm $i$, the consumer $i$ will obtain $t_i p_m z_m$ units of income. Therefore, the equilibrium income allocation among $N$ consumers can be written in the form:

$$\left(t_1 p_m z_m, t_2 p_m z_m, \ldots, t_N p_m z_m\right). \qquad (6)$$

---

[5] Then $y_1^* = t_1 z$, $y_2^* = t_2 z$ and so on.

If we denote by $\Pi = p_m z_m$ the total income (or GDP), the equilibrium income allocations (6) can be directly written as:

$$(R_1, R_2, ..., R_N), \tag{7}$$

where $R_i$ denotes the income of the $i$ th consumer and by (4) satisfies:

$$\begin{cases} R_i \geq 0 \quad for \quad i = 1, 2, ..., N \\ \sum_{i=1}^{N} R_i = \Pi \end{cases} \tag{8}$$

Similar to the statistical physics in which the microstate (1) is replaced by the energy allocation among particles, the general equilibrium (3) has been replaced by the income allocation among consumers, (7). Therefore, we can immediately introduce the concept of ensemble of economic system as follow:

(g). The ensemble of economic system consists of all possible equilibrium income allocations $(R_1, R_2, ..., R_N)$ satisfying (8).

Undoubtedly, if we apply the principle of maximum entropy, then we can easily obtain the most probable income distribution. However, to apply this principle, we must assume that every income allocation (or general equilibrium) $(R_1, R_2, ..., R_N)$ will occur with an equal probability. Next we show that the hypothesis of equal probability can be applied in a just society which captures Rawls' fairness. To see this, let us first introduce Arrow's Impossibility Theorem (AIT).

## 4. Social Choices

To help readers to understand AIT more easier, we consider a simple society in which there are four equilibrium income allocations (or general equilibria) only: $A_1$, $A_2$, $A_3$ and $A_4$. The main purpose of AIT is to answer the question about social choice: Which of these four income allocations is best for society. To accomplish this task, one may denote by $A = \{A_1, A_2, A_3, A_4\}$ the ensemble. Then, if one can specify some ranking of the income allocations in $A$ that reflects 'society's' preferences, one would find the best social choice. Unfortunately, AIT has refuted the existence of such 'society's' preferences [18]. Hence social members are not able to compare any two alternatives in $A$ from a point of view which is individually consistent and social consistent; otherwise, the social choice will be unfair. Now that comparing any two income allocations is unavailable, we might as well admit the indifference between all these income allocations; that is,

$$A_1 \sim A_2 \sim A_3 \sim A_4, \tag{9}$$

where, the symbol $\sim$ stands for the indifference relation.

Our treatment above agrees with Leibniz's *principle of the identity of indiscernibles* [19]. Since social members are indifferent between all equilibrium income allocations, they cannot ensure which equilibrium income allocation will be selected as a collective decision. This means that collective choices should be completely random. Although random choices occur, we can still ensure the probability of selecting each equilibrium allocation in a just society. To this end, we must apply Rawls' principle of social fairness; that is, Rawls' principle of fair equality of opportunity, a central topic in the theory of social justice. The principle holds that social and economic inequalities are to be arranged so that they are open to all under conditions of fair equality of opportunity, and the principle has radical implications for the design of social policy and legislation in modern democracies. Fair equality of opportunity can be regarded as an extension of the ideal of nondiscrimination, showing that the door of opportunity opens to all the members at least within the public realm of a civil society.

On the basis of pure procedural justice, Rawls argued that a just procedure would translate its fairness to the (allocation) outcomes [20]; therefore, a just society implies that each allocation outcome should be selected with equal opportunities. Applying Rawls' principle of fair equality of opportunity into (9) we immediately arrive at the hypothesis of equal probability:

$$P[A_1] = P[A_2] = P[A_3] = P[A_4] = \frac{1}{4}, \tag{10}$$

where, we denote by $P[X]$ the probability that the income allocation $X$ occurs (or is selected).

Obviously, (10) can be easily extended to the case where general equilibria (or income allocations) (7) are taken into account. Therefore, we can make an axiom about "absolute fairness" as below [13]:

**Axiom 1** (Rawls' Fairness)**:** If a competitive economy produces $\omega$ equilibrium outcomes, and if this economy is absolutely fair, then each equilibrium outcome will occur with an equal probability $\frac{1}{\omega}$.

Similar to the macrostate in statistical physics, we might use a set of non-negative numbers, $\{a_k\}_{k=1}^n = \{a_1, a_2, ..., a_n\}$, to denote a possible income distribution, where each $a_k$ represents that there are $a_k$ consumers each of who obtains $\varepsilon_k$ units of income[6], and that these $a_k$ consumers are distributed among $g_k$ industries. If one denotes by $\Omega(\{a_k\}_{k=1}^n)$ the number of income allocations that the income distribution $\{a_k\}_{k=1}^n$ contains, then one has [7,13]:

---

[6] Here $0 \leq \varepsilon_1 < \varepsilon_2 < ... < \varepsilon_n$.

$$\Omega\left(\{a_k\}_{k=1}^{n}\right) = \begin{cases} \prod_{k=1}^{n} \dfrac{(a_k + g_k - 1)!}{a_k!(g_k - 1)!} & (perfect\ competition) \\ \dfrac{N!}{\prod_{k=1}^{n} a_k!} \prod_{k=1}^{n} g_k^{a_k} & (monopolistic\ competition) \end{cases}, \quad (11)$$

where, perfect competition indicates the extreme (white-hot) competitions between homogeneous consumers (or firms), and monopolistic competition indicates the moderate competitions between heterogeneous consumers (or firms) [7,13].

## 5. Results

By the Axiom 1 we can apply the principle of maximum entropy into (11). Before proceeding to do this, it is here worth mentioning that $g_k$ may not be a large number (for example, single industry assumed by us implies $g_k = 1$ for $k = 1, 2, \ldots, n$); therefore, we must carefully derive the most probable income distribution for perfect competition, $\{a_k^*\}_{k=1}^{n}$.

Undoubtedly, if the Axiom 1 holds, $\{a_k^*\}_{k=1}^{n}$ must maximize $\ln \Omega\left(\{a_k\}_{k=1}^{n}\right)$, which for perfect competition can be written as:

$$\ln \Omega\left(\{a_k\}_{k=1}^{n}\right) = \sum_{k=1}^{n} \ln(a_k + g_k - 1)! - \sum_{k=1}^{n} \ln a_k! - \sum_{k=1}^{n} \ln(g_k - 1)!. \quad (12)$$

Because $a_k$ is a large number, by the Stirling formula

$$\ln m! \approx m(\ln m - 1) \text{ for } m \gg 1$$

the equation (12) can be rewritten in the form:

$$\ln \Omega\left(\{a_k\}_{k=1}^{n}\right) = \sum_{k=1}^{n} \left[(a_k + g_k - 1)\ln(a_k + g_k - 1) - a_k \ln a_k - g_k - \ln(g_k - 1)! + 1\right]. \quad (13)$$

Then by the method of Lagrange multipliers we have:

$$\ln\left(\frac{a_k^* + g_k - 1}{a_k^*} - \alpha - \beta \varepsilon_k\right) = 0 \quad (14)$$

for $k = 1, 2, \ldots, n$.

Thus, one obtains the most probable income distribution for perfect competition[7]:

---

[7] If $g_k \gg 1$, then one has $g_k - 1 \approx g_k$.

$$a_k^* = \frac{g_k - 1}{e^{\alpha + \beta \varepsilon_k} - 1} \tag{15}$$

By the same method in [7] we can finally get:

$$a_k^*(I) = \frac{g_k - I}{e^{\frac{\varepsilon_k - \mu}{\lambda \theta}} - I} \quad \begin{cases} I = 1 & (perfect\ competition) \\ I = 0 & (monopolistic\ competition) \end{cases} \tag{16}$$

for $k = 1, 2, \ldots, n$,

where, $\mu$ and $\theta$ denote the marginal labor-capital return and the marginal technology return of an economy respectively, and $\lambda$ is a positive constant.

For the case of multiple industries, by (16) the perfect competition may lead to an unstable economy (or economic crises) [7]. If there is only one industry in the economy, then we have the following proposition which will reproduce the standard result for perfect competition in neoclassical economics.

**Proposition 1:** If $g_k = 1$ for $k = 1, 2, \ldots, n$, then perfect competition implies that each firm (or consumer) occupies the same revenue (or income) level $\mu$; that is,

$$R_i = \mu \tag{17}$$

for $i = 1, 2, \ldots, N$.

*Proof.* Since $a_k^*(I = 1) = \frac{g_k - 1}{e^{\frac{\varepsilon_k - \mu}{\lambda \theta}} - 1}$ and $g_k = 1$ for $k = 1, 2, \ldots, n$, one will have

$a_k^*(I = 1) = 0$ for $k = 1, 2, \ldots, l-1, l+1, \ldots, n$, where $\varepsilon_l = \mu$. Then by $\sum_{k=1}^{n} a_k^*(I = 1) = N$, one immediately arrives at:

$$\begin{cases} a_l^*(I = 1) = N \\ a_k^*(I = 1) = 0 \quad for\ k = 1, 2, \ldots l-1, l+1, \ldots n \end{cases} \tag{18}$$

□

By the Proposition 1 we have shown that the method of statistical physics can reproduce the standard result of neoclassical economics. Even so, we still remind that the Proposition 1 only describes an idealized economy where a single industry is taken into account. However, if multiple industries are involved, then the inequality of revenue (or income) will occur [7]. It is here carefully noted that the "industry" in neoclassical economics has a rigid definition [13]: each industry corresponds to a different commodity[8]. For example, if there are only two types of commodities which are bread and clothes respectively in an "imaginary" country, then we say that there are two industries. Therefore, a single industry implies that the society only produces a type of commodity. This case almost never happens in a real society.

## 6. Discussion and Conclusion

(16) describes the most probable income distribution among social members (consumers) in the equilibrium economy. It is worth mentioning that (16) is due to the hypothesis of equal probability (Axiom 1) which arises because the society is assumed to be absolutely fair. However, human society *cannot* be absolutely fair, so (16) may be only suitable for a part of population. Indeed, Yakovenko's and Kusmartsev's empirical investigations have supported this result. Specifically, Yakovenko used the income data from 1983-2000 in USA to confirm that the majority of the population (lower class) obey Boltzmann distribution, see Figure 6 in Ref [6]; and Kusmartsev et al used the income data from 1996-2008 in USA to confirm that the majority of the population (lower class) obey Bose-Einstein distribution, see Figure 1 in Ref [8]. Moreover, Yakovenko's and Kusmartsev's investigations together agreed with that the small fraction of population (upper class) obey Pareto distribution. As is well known, Pareto distribution could be derived by using some unfair rules, e.g., the rule of "The rich get richer" in scale-free network [21]. This means that the upper class of population might involve unfair behaviors. By the same method proposed by Barabasi [21], Tao has got Pareto income distribution [22]:

$$a_k \propto \varepsilon_k^{-\gamma-1}, \qquad (19)$$

where, $k = 1, 2, \ldots, n$ and $\gamma \geq 1$.

In conclusion, Arrow-Debreu's general equilibrium equations in economics play the same role with many-body Newtonian equations in physics; therefore, each solution of the general equilibrium equations can be regarded as a possible microstate of the economic system. Then by Arrow's Impossibility Theorem and Rawls' principle of social fairness, the hypothesis of equal probability (Axiom 1) should be valid for an absolutely fair society as well. So by the principle of maximum entropy, the income distribution among social members will obey the spontaneous order[9], which emerges as Boltzmann distribution for a moderately competitive society and as Bose-Einstein distribution for an extremely competitive society. This means that the rule of income distribution in a just society is not the result of any process of collective choice (unlike expected by many welfare economists), but is an unplanned and spontaneous consequence (as expected by Hayek). Although the human society cannot be absolutely fair, the empirical investigations show that the majority of population (lower class) still well obeys the spontaneous order. However, the small fraction of population (upper class) will obey Pareto distribution which can be derived by using some unfair behavior rules.

---

[8] Adopting technical term in Arrow-Debreu's general equilibrium model, production vector's positive component denotes output and negative component denotes input. So the number of positive component stands for number of industries.

[9] According to Hayek's thought, a striking feature of the spontaneous economic order is that it is more likely to emerge or more able to survive than other economic orders [13]. Following this thought, with each economic order we can associate a possible individuals' income distribution. Accordingly, the income distribution with the largest probability will be naturally thought of as the spontaneous economic order.

# References

[1]. Aoyama, H., Yoshikawa, H., Iyetomi, H., Fujiwara, Y. (2010): Productivity dispersion: facts, theory, and implications. Journal of Economic Interaction and Coordination 5: 27-54

[2]. Foley, D. (1994): A statistical equilibrium theory of markets. Journal of Economic Theory 62: 321-345

[3]. Stanley, H. E., *et al.* (1996): Anomalous fluctuations in the dynamics of complex systems: from DNA and physiology to econophysics. Physica A 224: 302–321.

[4]. Dregulescu, A. A. and Yakovenko, V. M. (2000): Statistical mechanics of money. Eur. Phys. J. B 17: 723–729.

[5]. Dregulescu, A. A. and Yakovenko, V. M. (2001): Evidence for the exponential distribution of income in the USA. Eur. Phys. J. B 20: 585–589.

[6]. Yakovenko, V. M. and Rosser, J. B. (2009): Statistical mechanics of money, wealth, and income. Reviews of Modern Physics 81: 1703-1725

[7]. Tao, Y. (2010): Competitive market for multiple firms and economic crisis. Phys. Rev. E. 82: 036118

[8]. Kusmartsev, F. V. (2011): Statistical mechanics of economics I. Phys. Lett. A 375: 966.

[9]. Kürten, K. E. and Kusmartsev, F. V. (2011): Bose-Einstein distribution of money in a free-market economy. II. Europhysics Letters 93: 28003.

[10]. Tao, Y., Chen, X. (2012): Statistical physics of economic systems: a survey for open economies. Chinese Physics Letters 29: 058901

[11]. Flomenbom, O. (2011): Fairness in society, arXiv: 1112.0262

[12]. Chakraborti, A. et al (2014): Statistical Mechanics of Competitive Resource Allocation using Agent-based Models, arXiv:1305.2121

[13]. Tao, Y. (2013): Spontaneous Economic Order. arXiv:1210.0898

[14]. Mas-Collel, A., Whinston, M. D., Green, J. R. (1995): Microeconomic Theory. Oxford University Press. Page 579.

[15]. Varian, H. R. (1992): Microeconomic Analysis (Third Edition), Norton & Company, Inc. New York. Page 3.

[16]. Acemoglu, D. and Robinson, J. (2012): Why Nations Fail: The Origins of Power, Prosperity, and Poverty, Crown Business

[17]. Varian, H. R. (2003): Intermediate Microeconomics: A modern approach (Sixth Edition). Norton, New York. Page 340.

[18]. Jehle, G. A. and Reny, P. J. (2001): Advanced Microeconomic Theory (Second Edition). ADDISON-WESLEY. Page 243.

[19]. Arrow, K. J. (1963): Social choice and individual values. John Wiley& Sons, Inc., New York. Page 109.

[20]. Rawls, J. (1999): A Theory of Justice (Revised Edition), Cambridge, MA: Harvard University Press. Page 74.

[21]. Barabasi, A. L. and Albert, R., (1999): Emergence of scaling in random networks. Science 286 (5439): 509-512

[22]. Tao, Y. (2014): Theoretical research of law of society's income distribution-Based on Arrow-Debreu's general equilibrium theory. Doctoral Dissertation of Chongqing University.